\documentclass[10pt, conference, compsocconf]{IEEEtran}

\usepackage[cmex10]{amsmath}
\usepackage{cite}
\usepackage[pdftex]{graphicx}
\usepackage{algorithmic}
\usepackage{array}
\usepackage{tabularx}
\usepackage{multirow}
\usepackage{caption}
\usepackage{subcaption}
\usepackage{listings}

\begin{document}
%
\title{Sparse Allreduce: Efficient Scalable Communication for Power-Law Data}


\author{\IEEEauthorblockN{Huasha Zhao}
\IEEEauthorblockA{Computer Science Division\\
University of California\\
Berkeley, CA 94720\\
hzhao@cs.berkeley.edu}
\and
\IEEEauthorblockN{John Canny}
\IEEEauthorblockA{Computer Science Division\\
University of California\\
Berkeley, CA 94720\\
jfc@cs.berkeley.edu}
}

\maketitle

\begin{abstract}
Many large datasets exhibit power-law statistics: The web graph,
social networks, text data, clickthrough data etc. Their adjacency
graphs are termed {\em natural graphs}, and are known to be difficult
to partition. As a consequence most distributed algorithms on these
graphs are communication-intensive.  Many algorithms on natural graphs
involve an {\em Allreduce:} a sum or average of partitioned data which
is then shared back to the cluster nodes. Examples include PageRank,
spectral partitioning, and many machine learning algorithms including
regression, factor (topic) models, and clustering.  In this paper we
describe an efficient and scalable {\em Allreduce} primitive for
power-law data. We point out scaling problems with existing butterfly
and round-robin networks for Sparse Allreduce, and show that a hybrid
approach improves on both. Furthermore, we show that Sparse Allreduce
stages should be nested instead of cascaded (as in the dense case).
And that the optimum throughput Allreduce network should be a
butterfly of {\em heterogeneous degree} where degree decreases with
depth into the network. Finally, a simple replication scheme is
introduced to deal with node failures. We present experiments showing
significant improvements over existing systems such as PowerGraph and
Hadoop.
\end{abstract}


%
\IEEEpeerreviewmaketitle
\section{Introduction}
Power-law statistics are the norm for most behavioural datasets,
i.e. data generated by people, including the web graph, social
networks, text data, clickthrough data etc. By power-law, we mean that
the probability distributions of elements in one or both (row and
column) dimensions of these matrices follow a function of the form
\begin{equation}
p \propto d^{-\alpha}
\end{equation}
where $d$ is the degree of that feature (the number of non-zeros in
the corresponding row or column).  These datasets are large: 40 billion vertices for the web
graph, terabytes for social media logs and news archives, and
petabytes for large portal logs. Many groups are developing tools to
analyze these datasets on clusters \cite{low2010graphlab,
  gonzalez12powergraph, malewicz2010pregel, ZC13, kang2009pegasus,
  isard2007dryad, ranu2009graphsig, chent2007gapprox,
  yan2002gspan,papadimitriou2008disco}. While cluster approaches
have produced useful speedups, they have generally not leveraged
single-machine performance either through CPU-accelerated libraries
(such as Intel MKL) or using GPUs. Recent work has shown that very
large speedups are possible on single nodes \cite{CZKDD13,ZC13}, and
in fact for many common machine learning problems single node
benchmarks now dominate the cluster benchmarks that have appeared in
the literature \cite{CZKDD13}.

Its natural to ask if we can further scale single-node performance on
clusters of full-accelerated nodes. However, this requires
proportional improvements in network primitives if the network is not
to be a bottleneck.  In this work
we are looking to obtain one to two orders of magnitude improvement in
the throughput of the Allreduce operation.

Allreduce is a rather general primitive that is integral to many
distributed graph mining and machine learning algorithms. In an
Allreduce, data from each node, which can be represented as a vector
$v_i$ for node $i$, is reduced in some fashion (say via a sum) to
produce an aggregate
$$v = \sum_{i=1,\ldots,M} v_i$$
and this aggregate is then shared across all the nodes. 

In many applications, and in particular when the shared data is large,
the vectors $v_i$ are sparse. And furthermore, each cluster node may
not require all of the sum $v$ but only a sparse subset of it. 
We call a primitive which provides this capability a {\em Sparse
Allreduce}. By communicating only those values that are needed by
the nodes Sparse Allreduce can achieve orders-of-magnitude speedups
over dense approaches. 

The aim of this paper is to develop a general Sparse Allreduce
primitive, and tune it to be as efficient as possible for a given
problem. We next show how Sparse Allreduce naturally arises in
algorithms such as PageRank, Spectral Clustering, Diameter Estimation,
and machine learning algorithms that train on blocks (mini-batches) of
data, e.g. those that use Stochastic Gradient Descent(SGD) or Gibbs
samplers.

\subsection{Applications}
\label{sec:Applications}

\subsubsection{MiniBatch Machine Learning Algortihms}
\begin{figure}
\centering
\includegraphics[scale = 0.5]{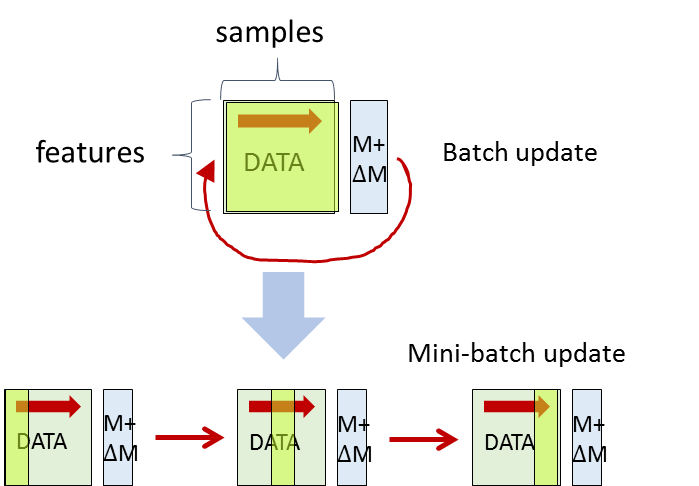}
\caption{Batch update vs. mini-batch update}
\label{minibatch}
\vspace*{-0.2in}
\end{figure}

Recently there
has been considerable progress in sub-gradient algorithms
\cite{le2004large,duchi2011adaptive} which partition a large dataset into
mini-batches and update the model using sub-gradients, illustrated in
Figure \ref{minibatch}. Such models achieve many model updates in a single
pass over the dataset, and several benchmarks on large datasets show
convergence in a single pass \cite{le2004large}. While sub-gradient
algorithms have relatively slow theoretical convergence, in practice
they often reach a desired loss level much sooner than other
methods for problems including Regression, Support Vector Machines,
factor models, and several others. 

Finally, MCMC algorithms such as Gibbs samplers involve updates to a
model on every sample.  In practice to reduce communication overhead,
the sample updates are batched in very similar fashion to sub-gradient
updates \cite{smola2010architecture}.

All these algorithms have a common property in terms of the input
mini-batch: if the mini-batch involves a subset of features 
$\{f_1,\ldots,f_n\}$, then a gradient update commonly uses input only from,
and only makes updates to,  the subset of the model that is
projected onto those features. This is easily seen for factor and regression
models whose loss function has the form 
$$l = f(AX)$$
where $X$ is the input mini-batch, A is a matrix which partly parametrizes the model, 
and f is in general a non-linear function. The derivative of loss, which 
defines the SGD update, has the form
$$dl/dA = f^{\prime}(AX) X^T$$
from which we can see that the update is a scaled copy of $X$,
and therefore involves the same non-zero features. 

\subsubsection{Iterative Matrix Product}
Many graph mining algorithms use repeated
matrix-matrix/vector multiplication. Here are some representative examples.
\medskip

\noindent{\bf PageRank}
Given the adjacency matrix $G$ of a graph on $n$ vertices with
normalized columns (to sum to 1), and $P$ a vector of vertex scores,
the PageRank iteration in matrix form is:
\begin{equation}
P^\prime = \frac{1}{n} + \frac{n-1}{n}G P
\end{equation}

\noindent{\bf Diameter Estimation}
In the HADI \cite{kang2008hadi} algorithm for diameter estimation, the number of neighbours within hop $h$ is encoded in a probabilistic bit-string vector $b^h$. The vector is updated as follows:
\begin{align}
b^{h+1} = G \times_{or} b^h.
\end{align}
Again $G$ is the adjacency matrix and operation $\times_{or}$ replaces addition in matrix-vector product is replaced by bitwise OR operation. 
\medskip

\noindent{\bf Spectral Graph Algorithms}
Spectral methods make use of the eigen-spectrum (some leading set of
eigenvalues and eigenvectors) of the graph adjacency matrix. Almost all eigenvalue
algorithms use repeated matrix-vector products with the matrix. 

To present one of these examples in a bit more detail:
PageRank provides an ideal motivation for Sparse Allreduce. The
dominant step is computing the matrix-vector product $G~ P$. We assume
that edges of adjacency matrix $G$ are distributed across machines
with $G_i$ being the share on machine $i$, and that vertices $P$ are
also distributed (usually redundantly) across machines as $P_i$ on
machine $i$.  At each iteration, every machine first acquires a sparse {\em
  input} subset $P_i$ corresponding to non-zero columns of its share
$G_i$ - for a sparse graph such as a web graph this will be a small
fraction of all the columns.  It then computes the product $Q_i = G_i
P_i$. This product vector is also sparse, and its nonzeros correspond
to non-zero {\em rows} of $G_i$. The input vertices $P_i$ and the
output vertices $Q_i$ are passed to a sparse (sum) Allreduce, and the
result loaded into the vectors $P'_i$ on the next iteration will be
the appropriate share of the matrix product $G P$. Thus a requirement
for Sparse Allreduce is that we be able to specify a vertex subset
going in, and a different vertex set going out (i.e. whose values are
to be computed and returned). 
\medskip

\subsection{Summary of Work and Contributions}

In this paper, we describe Sparse Allreduce, a communication primitive
that supports high performance distributed machine learning on sparse
data. Our Sparse Allreduce has the following properties:
\begin{enumerate}
\item Each network node specifies a sparse vector of output values, and
a vector of input {\em indices} whose values it wants to obtain from
the protocol.

\item Index calculations (configuration) can be separated from value 
calculations and only computed once for problems where the indices
are fixed (e.g. Pagerank iterations). 

\item The Sparse Allreduce network is a {\em nested, heterogeneous} butterfly. 
By heterogeneous we mean that the butterfly degree $k$ differs from one
layer of the network to another. By nested, we mean that values pass ``down''
through the network to implement an scatter-reduce, and then back up 
through the same nodes to implement an allgather. 

\end{enumerate}


Sparse Allreduce is modular and easy to run, and requires only a
mapping from node ids to IP addresses. Our current implementation is in
pure Java, making it easy to integrate with Java-based cluster
systems like Hadoop, HDFS, Spark, etc.

\begin{figure*}[t!]
\centering
\includegraphics[scale = 0.35]{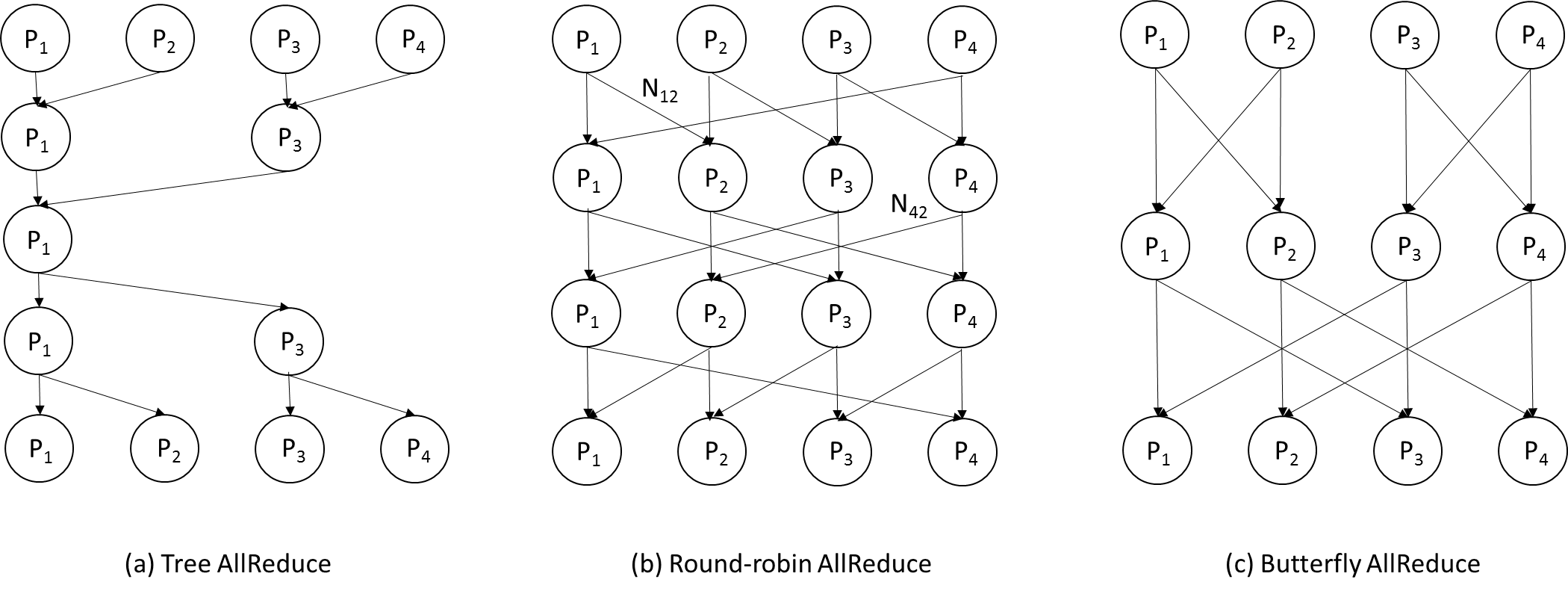}
\caption{Allreduce Topologies} \label{network}
\end{figure*}

The key contributions of this paper are the following:

\begin{itemize}
\item Sparse Allreduce, a programming primitive that supports
  efficient parallelization of a wide range of iterative
  algorithms, including PageRank, diameter estimation, and mini-batch
  gradient algorithms for Machine Learning, and others. 

\item A number of experiments on large datasets with billions of
  edges. Experimental results suggest that Sparse Allreduce
  significantly improves over prior work, by factors of 5-30.

\item A replication scheme that provides a high-degree of fault-tolerance
 with modest overhead. We demonstrate that Allreduce with our replication
scheme can tolerate about $\sqrt{M}$ node failures on an $M$-node network,
and that node failures themselves do not slow down the operation.

\end{itemize}

The rest of the paper is organized as follows. Section II reviews
existing Allreduce primitives, and highlights their difficulties when
applied to large, sparse datasets.  Section III introduces Sparse
Allreduce, its essential features, and an example network.
Section IV and V describe its optimized implementation and fault tolerance respectively. Experimental results are presented in
Section VI. We summarize related works in Section VII, and finally Section
VIII concludes the paper.

\section{Background: Allreduce on Clusters}
Allreduce is a common operation to aggregate local model update in distributed machine learnings. This section reviews the practice of data partition across processors, and popular Allreduce implementations and their limitations.

\subsection{AllReduce}
When data is partitioned across processors, local updates must then be combined by an additive or average reduction, and then
redistributed to the hosts. This process is commonly known as Allreduce. Allreduce is commonly implemented with 1) tree structure\cite{langford2007vowpal}, 2) simple round-robin in full-mesh networks or 3) butterfly topologies\cite{patarasuk2009bandwidth}.

\subsubsection{Tree Reduce}
The tree reduce topology is illustrated in Figure
\ref{network}(a). The topology uses the lowest overall bandwidth for
atomic messages, although it effectively maximizes latency since the
delay is set by the slowest path in the tree. It is a reasonable
solution for small, dense (fixed-size) messages.  A serious limitation
for Sparse Allreduce applications is that the entire sum is held and
distributed by the bottom half of the network. i.e.  the length of the
sparse sums is increasing as one goes down the layers of this network,
and eventually encompasses the entire sum. Thus the time to compute
sums is increasing layer-by-layer and one effectively loses the
advantage of parallelism. It is not practical for the problems of
interest to us, and we will not discuss it further.

\subsubsection{Round-Robin Reduce}
In round-robin reduce, each processor communicate with all other
processors in a circular order, as presented in Figure
\ref{network}(b). Round-robin reduce achieves asymptotically optimal
bandwidth, and optimal latency {\em when packets are sufficiently large}
to mask setup-teardown times. In practice though, this requirement is often
not satisfied, and there is no way to tune the network to avoid this problem. 
Also, the very large (quadratic in M) number of messages make this network
more prone to failures due to packet corruption, and sensitive to latency outliers. 

\begin{figure}[t!]
\centering
\includegraphics[scale = 0.4]{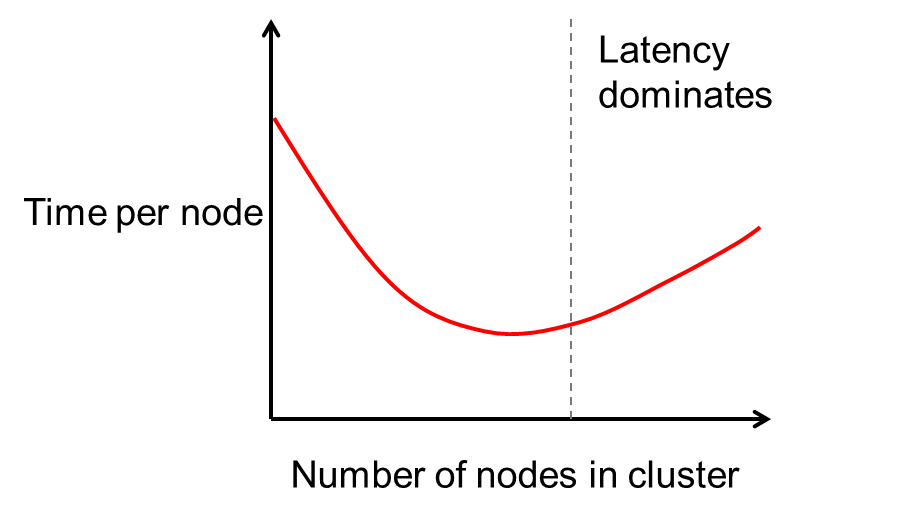}
\caption{Scalability of round-robin network of 64 Amazon EC2 nodes}
\label{throughput}
\vspace*{-0.2in}
\end{figure}

In our experiment setup of a 64-node Amazon EC2 cluster with 10Gb/s
inter-connect, the optimal packets size is 1M-10M to mask message
sending overhead. As illustrated in Figure \ref{throughput}, for
smaller packets, latency dominates the communication process, so the runtime per node will goes up when distributing data to larger clusters. In many problems of
interest, and e.g. the Twitter follower's graph and Yahoo's web graph,
the packet size in each round of communication under a round-robin
network is much smaller than optimal. This causes significant
inefficiencies.

\subsubsection{Butterfly Network}
In a butterfly network, every node computes some function of the
values from its {\em in} neighbours (including its own) and outputs to
its {\em out} neighbours. In the binary case, the neighbours at layer
$d$ lie on the edges of hypercube in dimension $d$ with nodes as
vertices.  The cardinality of the neighbour set is called the degree
of that layer. Figure \ref{network} demonstrates a $2 \times
2$ butterfly network and Figure \ref{butterfly} shows a $3 \times
2$ network. A binary butterfly gives the lowest latency for Allreduce
operations when messages have fixed cost.

Faults in the basic butterfly network affect the outputs, but on a
subset of the nodes. Simple recovery strategies (fail-over to the
sibling just averaged with) can produce complete recovery since every
value is computed at two nodes. However, butterfly networks involve
higher bandwidth.

While these networks are ``optimal'' in various ways for dense
data, they have the problems listed above for sparse data. However,
we can hybridize butterfly and round-robin in a way that gives us
the good properties of each. Our solution is a $d$-layer butterfly
where each layer has degree $k_1,\ldots,k_d$. Communication within
each group of $k_i$ will use the Allreduce pattern. We adjust $k_i$
for each layer to the largest value that avoids saturation (packet
sizes below the practical minimum discussed earlier). Because the
sum of message lengths decreases as we do down layers of the network,
the optimal $k$-values will also typically decrease.

\subsection{Partitions of Power-Law Data}

As shown in \cite{gonzalez12powergraph}, edge partitioning is much more effective
for large, power-law datasets than vertex partitioning. The paper \cite{gonzalez12powergraph}
describes two edge partitioning schemes, one random and one greedy. Here we will only
use random edge partitioning - we feel this is more typically the case for data that is ``sitting
in the network'' although results should be similar for other partitioning schemes.

\section{Sparse Allreduce}


In this section, we describe a typical work flow of distributed
machine learning, and introduce Sparse Allreduce. Example usage of
Sparse Allreduce is discussed in Section III-B.

\subsection{Overview of Sparse AllReduce}
A typical distributed learning task starts with graph/data partitioning,
followed by a sequence of alternating model update and model
Allreduce. The ``data'' may directly represent a graph, or may
be a data matrix whose adjacency graph is the structure to be partitioned. 
Canonical examples are PageRank and other matrix power calculations, or
large-model machine learning algorithms such as LDA. 

To avoid clustering of high-degree vertices with similar indices, we
first apply a random hash to the vertex indices (which will effect a
random permutation). We then sort and thereafter maintain indices in
sorted order - this is part of the data structure creation and we
assume it is done before everything else.

Then the vertex set in a group of $k$ nodes is split into $k$
ranges. Because of the initial index permutation, we are effectively
partitioning vertices into $k$ sets randomly, but it is much more
efficient to do using the sorted index sets. The partitioning is
literally splitting the data into contiguous intervals as show in
figure \ref{butterfly}, using a linear-time, memory-streaming
operation. Each range is sent to one of the neighbours (or to self) in the
group.

Each node in the layer below receives sparse vectors in one of the
sub-ranges and sums them. For performance reasons, we implement the
sums of $k$ vectors using a tree - direct addition of vectors to a
cumulative sum has quadratic complexity. Hashing has very bad memory
coherence and is about an order of magnitude slower than coherent
addition of sorted indices. For the tree addition, the input vectors
form the leaves of the tree. The leaves are grouped together in pairs
to form parent nodes, and each parent nodes holds a sum of its
children. We continue summing siblings up to a root node.  This
approach has $O(N log k)$ complexity ($N$ is total length of all
vectors) if there were no index collisions. But thanks to the
high frequency of such collisions for power-law data, the total lengths
of vectors decreases as we go up the tree. This is bounded by a 
multiplicative factor less than one, so the practical complexity
is $O(N)$ for this stage. In terms of constants, it was about 5x faster
overall than a hash table approach. 

This stage also produces a very helpful
compression of the data: i.e. many indices of input vertices collide,
and the total length of all vectors across the cluster at the second
layer is a fraction of the amount at the first layer.

The same process is repeated at the layer below, and continues
until we reach the bottom layer of the network. At this point, we
will have the sum of all input vectors, and it will be split into
narrow and distinct sub-ranges representing $R/M$ where $R$ is the
original range of indices, and $M$ is the number of machines. 

From here on, the algorithm involves only distribution of results
(allgather). Each layer passes up the values that were requested by a
parent (and whose indices were saved during the configuration step) to
that specific parent. The indices of those values are sorted and they
lie in distinct ranges, and the parent has only to concatenate them to
produce its final output sparse vector.


\subsection{Use Case Examples}
We provide two methods \verb config  and
  \verb reduce , for the programmers. Configuration involves passing down
    the outbound indices ( an array of vertex indices to be reduced
    (outbound) and inbound indices (an array of indices to
    collect). After configuration, the reduce function is called
    to obtain the vertex values for the next iteration. The
    reduce function takes in the vertex values to be reduced
    (corresponding to outbound vertices) and returns the vertex values
    for the next update (corresponding to inbound vertices). The
    following code examples show the usage of our primitive to run the
    PageRank algorithm and mini-batch update algorithm.\\

{\bf PageRank}:
{\tt \small
\begin{lstlisting}[frame=single]  
var out = outbound(G)
var in = inbound(G)
config(out.indices, in.indices)
for(i <-0 until iter){
  in.values = reduce(out.values)
  out.values = matrix_vec_multi(G,in.values)
}
\end{lstlisting}
}
In PageRank, the graph is static, so only one call of config is required at the beginning of iterations until convergence.\\

{\bf Mini-Batch Algorithm}: 
{\tt \small
\begin{lstlisting}[frame=single]  
for(i <-0 until iter){
  var Di = D(i*b until (i+1)*b)
  var out = outbound(Di)
  var in = inbound(Di)
  config(out.indices, in.indices)
  in.values = reduce(out.values)
  out.values = model_update(Di,in.values)
}
\end{lstlisting}
}
At the beginning of each iteration, a mini-batch of data of size $b$ is loaded to compute the new update. The graph of each mini-batch is dynamic, as a result, config need to be called before every reduce and model update.

\section{Implementation with a Butterfly of Heterogeneous Degrees}
\begin{figure*}[t!]
\centering
\includegraphics[scale = 0.25]{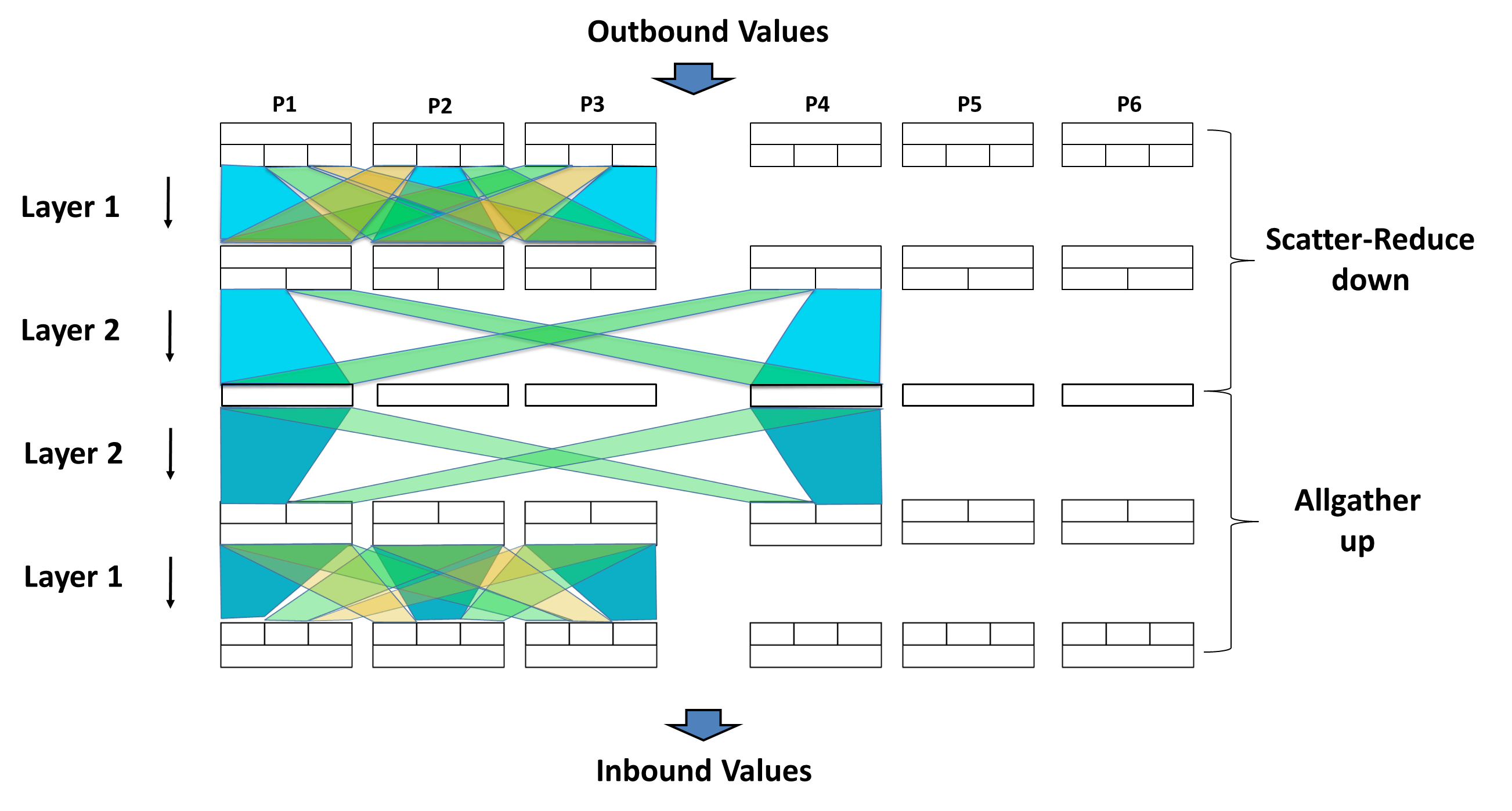}
\caption{Nested Sparse Allreduce within a heterogeneous-degree ($3\times2$) butterfly network .}
\label{butterfly}
\end{figure*}  
We have implemented Sparse Allreduce on a butterfly network with a layered design, as illustrated in Figure \ref{butterfly}. Each layer is characterized by two sets of neighbours the processor receive/send packets from/to and a set of indices/values to be exchanged. In a butterfly network, each processor sums vectors of indices from above, and partitions
and transmits the sum to below.


\subsection{Layered Config and Reduce}
Sparse Allreduce has two phases: config step and reduce step.  

In the config phase, each processor computes a map for each input vector. This map maps
indices from the input vector to the sparse sum of all input vectors. The maps
facilitate addition of values from above, and then the allgather stage going up. 

In the reduce phase, processors exchange the vertex values to be
reduced. Only vertex values are communicated, because vertex indices
are already hard-coded in the maps. 

We described here config and reduce methods separately, since this simplifies the
explanation and the code does include these operations. We also provide a combined
config-reduce method that performs both operations in a single round of communication
at each layer. i.e. the indices and values during the down phase are sent with the 
same messages. \\

%
{\bf Code for config and reduce:}
{\tt \small
\begin{lstlisting}[frame=single] 
//D: total levels of the butterfly
//setup indices
public void config(IVec downi, IVec upi) {
  IVec [] outputs = new IVec[2];
  for (int d = 0; d < D; i++) {
    layers[d].config(downi, upi, outputs);
    downi = outputs[0];
    upi = outputs[1];
  }
  finalMap = mapInds(upi, downi); 
}
//reduce values
public Vec reduce(Vec downv) {
  for (int d = 0; d < D; d++) {
    downv = layers[d].reduceDown(downv);
  }
  Vec upv = downv.mapFrom(finalMap); 
  for (int d = D-1; d >= 0; d--) {
    upv = layers[d].reduceUp(upv);
  }
  return upv;
}
\end{lstlisting}
}

Reduce layers are nested instead of cascaded as in the dense Allreduce
butterfly fly network. The reduce-scatter (down) computes the reduced
values and the allgather (up) redistributes the final values to
hosts. If we used a traditional layered butterfly without passing back
up through the same nodes, it would be necessary to push the inbound
indices along all the way through the network along with their values.
This would increase the overall size of configuration messages by about
50\%.

\subsection{Heterogeneous Butterfly Network}

In a heterogeneous butterfly network, the degree of {\em in} and {\em
  out} neighbours is different from layer to layer. Figure
\ref{butterfly} illustrates a $3 \times 2$ network, where each
processor talks to 3 neighbours in layer 1 and 2 in layer 2. The
benefit of using heterogeneous degree is work balance.

The heterogeneous butterfly network is a hybrid of pure round-robin
network and standard (degree 2) butterfly. In pure round-robin, packet
size in each round of communication is constrained to be $C/M^2$ where
C is the total dataset size, and M is number of machines. This may be too small - smaller than the
packet overhead. For example, in the Twitter followers' graph, the
packet size is around 0.5 MB in a 64 node round-robin network. Our
tests suggested the effective packet floor for these EC2 nodes is 2-4
MB.  On the other hand, in a degree 2 butterfly, the large number
of layers leads to much higher overall communication.

The heterogeneity of layer degrees allows us to tailor packet
size (which becomes $C/M/k$) with $k$. We can also deal with issues
like latency outliers across the network. Smaller $k$ values will
reduce the effects of latency outliers. 

Larger $k$ values are desirable, so long as they do not reduce message
sizes below the effective floor. Larger $k$ values leave less work to be done
in subsequent layers, and also reduce the total vector size in
the next layer because of index collisions. The more vectors that
are summed (and the number will be $k$) in the layer below, the more collisions of 
matching indices will occur, and each collision implies a reduction of
the number of indices below. 

Because of the reduction of total vector lengths in the layer below, the optimal
$k$ value will also be smaller (or we will hit the packet size floor again).

\subsection{Multi-Threading and Latency Hiding}
Scientific computing systems typically maintain a high degree of synchrony between
nodes running a calculation. In cluster environments, we have to allow for many
sources of variability in node timing, latency and throughput. While our network
conceptually uses synchronized messages to different destinations to avoid congestion,
in practice this does not give best performance. Instead we use multi-threading and
communicate opportunistically. i.e. we start threads to send all messages concurrently,
and spawn a thread to process each message that is received. In the case of replicated
messages, once the first message of a replicate group is received, the other threads
listening for duplicates are terminated and those copies discarded. 
Still, the network interface itself is a shared resource, so we have to be careful
that excessive threading does not hurt performance through switching of
the active message thread. The effects of varying the thread count is shown in Figure
\ref{threads}.

\subsection{Language and Networking Libraries}
Sparse Allreduce is currently implemented using standard Java
sockets. We explored several other options including OpenMPI-Java,
MPJexpress, and Java NIO.  Unfortunately the MPI implementations
lacked key features that we needed to support multi-threading,
asynchronous messaging, cancellation etc., or these features did not
work through the Java interface. Java NIO was simply more complex to
use without a performance advantage. All of the features we needed
were easily implemented with sockets, and ultimately they were a
better match for the level of abstraction and configurability that we
needed.

We acknowledge that the network interface could be considerably improved. 
The ideal choice would be RDMA over Ethernet (RoCE), and even better
RoCE directly from the GPUs. This feature in fact already exists (as
GPUdirect for NVIDIA CUDA GPUs). But it currently only available for 
infiniband networks. Other benchmarks of this technology suggest a
4- 5-fold improvement should be possible. 

\section{Fault Tolerance}
Machine failure is a reality of large clusters.  We introduce in this
section a simple but relatively efficient fault tolerance mechanism to
deal with multiple node failures.

\subsection{Data Duplication}
Our approach is to replicate by a replication factor $r$, the data on
each node, and all messages. Thus data on machine $i$ also appears on
the replicas $M+i$ through $i+(r-1)*M$. Similarly every config and
reduce message targeted at node $j$ is also sent to replicas $M+j$
through $j+(r-1)*M$. When receiving a message expected from node $j$,
the other replicas are also listened to.  The first message received
is used, and the other listeners are cancelled.

This protocol completes unless all the replicas in a group are dead.
e.g. when the replication factor is two, the probability of this
happening with random failures on an $M$-node network is about
$\sqrt{M}$ (birthday paradox).

\subsection{Packets Racing}
Replication by $r$ increases per-node communication by $r$ in the
worst case (cancellations will reduce it somewhat). There is some
performance loss because of this, as shown in the next section.
On the other hand, replication offers potential gains on networks with
high latency or throughput variance, because they create a race for
the fastest response (in contrast to the non-replicate network which 
is instead driven by the {\em slowest} path in the network. 

\section{Experiments}
\label{sec: perf}

In this section, we evaluate the performance and scalability of Sparse Allreduce, in comparison with other popular systems including Hadoop, Spark and PowerGraph. Three datasets are primarily studied in this section.
\begin{itemize}
\item Twitter follower's graph. The graph consists of 60 million vertices and 1.5 billion edges. Figure 1 in \cite{gonzalez12powergraph} shows the power-law property of this graph. 
\item Yahoo! Altavista web graph. This is one of the largest publicly available web graphs with 1.4 billion vertices and 6 billion edges.
\item Twitter document-term graph. The dataset consists of 1 billion unique tweets with 40 million uni-gram bag-of-words features each. The data is collected using Twitter API during the March of 2013, which provides 10\% sampled ``gardenhose'' twitter stream in the XML format.
\end{itemize}

All the experiments are conducted on the Amazon EC2 cluster consists of Cluster Compute nodes (cc1.4xlarge). Each node is equipped with 8 virtualized cores and they are interconnected by $10$Gb/s Ethernet. 

\subsection{Sparsity of the Datasets}
Table \ref{sparsity} demonstrates the sparsity of the partitioned
datasets. The Twitter followers' graph and Yahoo web graph are
partitioned across 64 processors using random edge partition, and the model
size is the total number of vertices in the graph. While the Twitter
document-term graph are partitioned by hour ot tweet; each partition is one
``mini-batch'' to feed into some sub-gradient/online method. The model
dimension is the number of uni-gram features.

\begin{table*}
\centering
\caption{Sparsity of the partitioned datasets}
\begin{tabular}{|c|c|c|c|} \hline
Data Set & Twitter follower's graph & Yahoo web graph & Twitter document-term graph\\ \hline
Partition \# of vertices & 12.1M &  48M & 5.1M\\
Total \# of vertices & 60M & 1.6B & 40M  \\ 
Percentage of total vertices & 0.21 & 0.03 & 0.12 \\
\hline \end{tabular}
\label{sparsity}
\end{table*}

As illustrated in the Table \ref{sparsity}, all dataset demonstrates
strong sparsity after partition. The Yahoo web graph is the biggest one in
terms of model size, and it is also the most sparse one among the
three, each partition only holds 3 percent of all the
vertices.

\subsection{Optimal System Parameters}

We described the trade-offs between round-robin and binary butterfly
earlier. In this section, we empirically determine the optimal
configuration of the butterfly degrees to deliver the best Sparse
Allreduce performance.

\begin{figure}
\centering
\includegraphics[scale = 0.5]{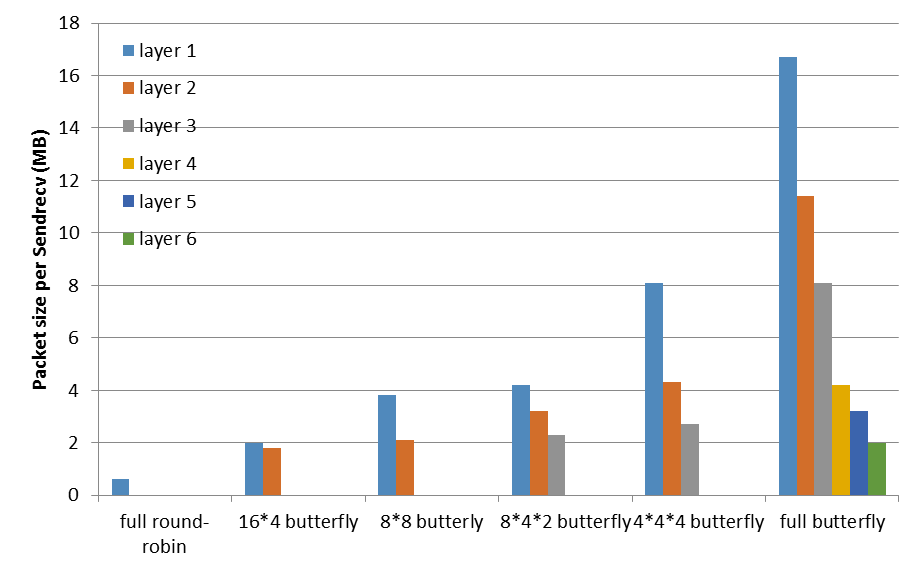}
\caption{Packet size at different level of butterfly network}
\label{packets}
\vspace*{-0.2in}
\end{figure}

Figure \ref{packets} plots the packet sizes at different level of butterfly network for different configurations in a 64-node cluster, which holds the random (edge) partitioned Twitter followers' graph. As illustrated from the figure, the 64 round-robin topology sends 0.5MB of packets each round which unlikely to fully utilize the bandwidth. Also for the butterfly configuration, although the packet size is decaying with depth into the network, the more layer we have, the more duplicated message we send. The full butterfly with degree 2 ends up sending packets of 17MB for each machine in the first round of communication.

\begin{figure*}
        \centering
        \begin{subfigure}[b]{0.45\textwidth}
		\centering
		\includegraphics[scale = 0.5]{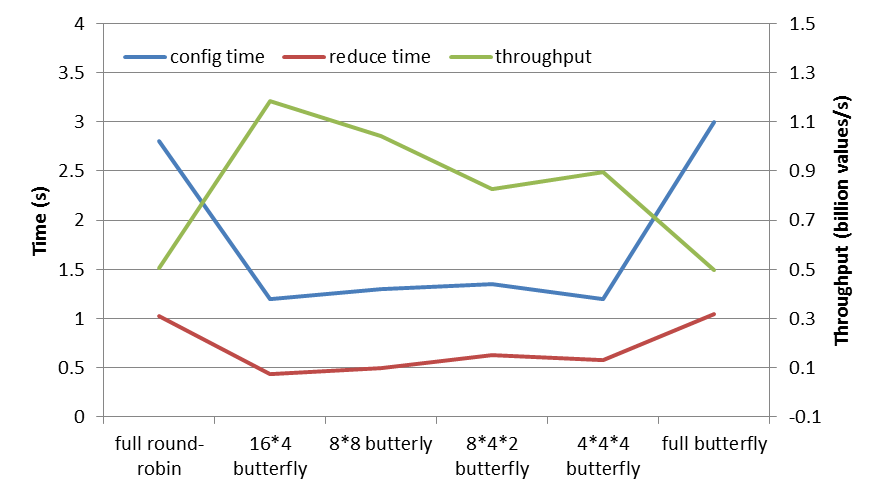}
		\caption{Twitter followers' graph}
		\label{strong}
        \end{subfigure}
        \begin{subfigure}[b]{0.45\textwidth}
		\centering
		\includegraphics[scale = 0.5]{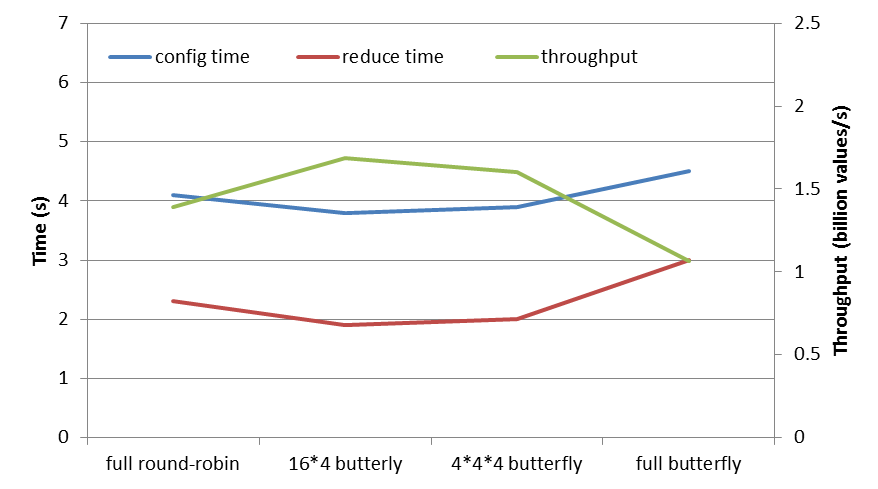}
		\caption{Yahoo web graph}
		\label{speedup}
        \end{subfigure}
        \caption{Allreduce time per iteration and throughput}
\label{comm}
\end{figure*}

Figure \ref{comm} plots the average reduce time per iteration and throughput for different configuration, for both Twitter followers' graph and Yahoo web graph. Throughput is measured in terms of total billions of input values reduced per second. From the figure, we can see the best configurations for both graph is $16 \times 4$. This has already been hinted in Figure \ref{packets}, for topology $16\times4$, communication is almost evenly distributed across two layers of the network; this balance prevents under-utilization of bandwidth by small packets.

It's not surprising to see that the round robin is closer to the optimal in the Web graph. The Web graph is much bigger in size, so latency is less of a problem when distributed to 64-node networks. However, round-robin may get into scaling issues when distributed to more machines.

\subsection{Effect of Multi-Threading}
\begin{figure}
\centering
\includegraphics[scale = 0.5]{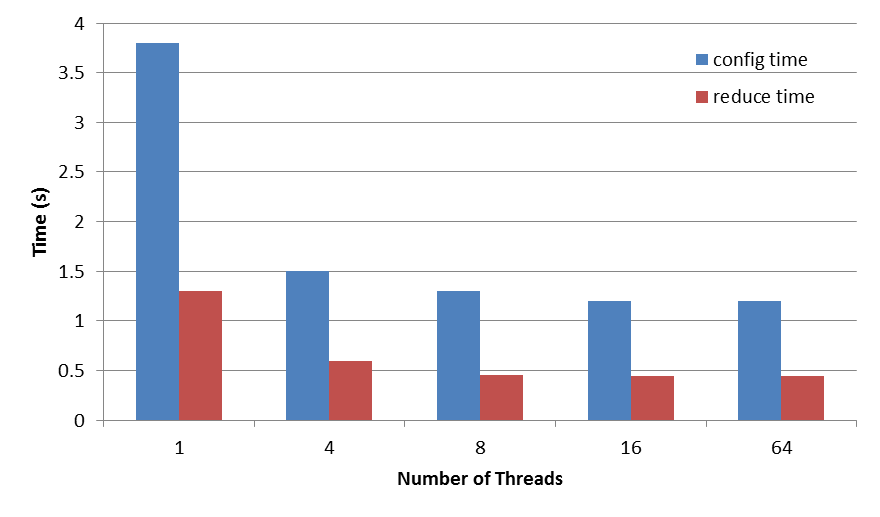}
\caption{Runtime comparison between different thread levels.}
\label{threads}
\vspace*{-0.2in}
\end{figure}

We compare the Allreduce runtime for different thread levels in Figure
\ref{threads}. All the results are run under the $16 \times 4$
configuration. Significant performance improvement can be observed by
increasing from single thread up to 4 threads, and it is also clear
from the figure that the benefit of adding thread level is marginal beyond 8 threads
(remember we are running on 8-core machines). However, there is no
penalty to add more threads: resources are just being shared among the
thread pool.

\subsection{Cost of Fault Tolerance}
Table \ref{deadnode} demonstrates the overhead of data replication in
terms of runtime. We compare a $8 \times 4$ network with replication
with a $16 \times 4$ network and a $8 \times 4$ network with no
replication. The $8 \times 4$ network with error-tolerance consumes
the same amount resource as the $16 \times 4$ network: both requires
64 machines. The data is partitioned into 32 pieces and each piece is
hosted by 2 machines. It doubles the resource requirement in
comparison with $8 \times 4$ network with no fault tolerance.

As shown in the table, the impact of data duplication on runtime is
moderate. In the $8 \times 4$ network, the replication version is only
10-15\% slower than the no replication version.  Given 64 machines, the
error-tolerance runs 50-60 \% slower than without
error-tolerance. 

\begin{table*}

\caption{Cost of Fault Tolerance}
\centering
\begin{tabular}[l]{|c|c|c|c|c|c|c|} 
\hline
\shortstack{System \\ Configuration} & \shortstack{$16 \times 4$ network \\replication=0} & \shortstack{$8 \times 4$ network \\replication=0} & \shortstack{$8 \times 4$ network\\ replication=1} & \shortstack{$8 \times 4$ network\\ replication=1} & \shortstack{$8 \times 4$ network \\replication=1} & \shortstack{$8 \times 4$ network\\ replication=1}\\ \hline
Number of dead nodes & 0 & 0 & 0 & 1 & 2 & 3 \\ \hline
Config time (s)& 1.2 & 1.3 & 1.51 & 1.49 & 1.52 & 1.51\\ \hline
Reduce time (s) & 0.44 & 0.60 & 0.75 & 0.73 & 0.76 & 0.74 \\  \hline 
\end{tabular}
\label{deadnode}
\end{table*}

We also compare the runtime for different number of node
failures. With no replication, the system cannot compute the correct
reduce results. The replicated version is able to compute the correct
result with node failures most of the time unless all the nodes in a
replication set are lost. For replication by two, the expected number
of failures to cause this is about $\sqrt{M}$ for $M$ total machines
(birthday paradox).

\subsection{Performance and Scalability}

In this section, we show the performance and scalability of Sparse Allreduce performance and scalability by running PageRank algorithm on clusters of different size and different systems. PageRank is implemented on top of BIDMat, an interactive matrix library written in Scala that fully utilize hardware accelerations (Intel MKL). So the computation is already an order of magnitude faster than pure Java. the Twitter follower's graph and Yahoo web graph. 

\begin{figure}
\centering
\includegraphics[scale = 0.5]{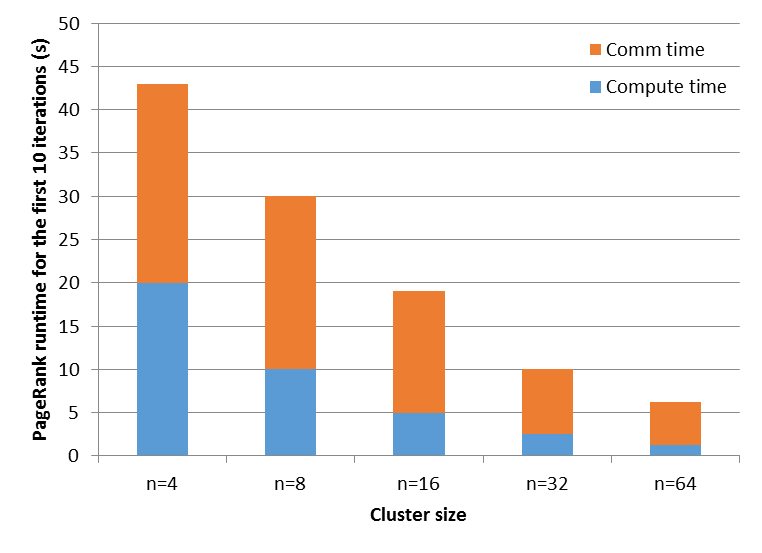}
\caption{Sparse Allreduce scaling and compute/comm break down.}
\label{scale}
\vspace*{-0.2in}v
\end{figure}

The scaling of Sparse Allreduce is illustrated in Figure \ref{scale}. The
figure plots the total runtime in the first 10 iteration against
cluster size. The configuration is optimally tuned individually for
different cluster size. We also present the runtime breakdown (into
computation and communication). As shown in the figure, the system
scales well up to 64 nodes. However, communication starts to dominate
the runtime for larger clusters. Particularly, for the 64 node
cluster, communication takes up to 80\% of overall runtime.

It is also worth pointing out that the overall achieved bandwidth is
around 2Gb/s on EC2 which is much smaller than the rated 10Gb/s of the
network.  This is not a bad number for the communication technology
used (pure Java sockets). Socket performance in HPC has been
extensively studied, and it is well-known that Java sockets achieve a
maximum of a few Gbits/sec. There are several technologies available
which would better this figure, however at this time there are
barriers to their use on commodity clusters. RDMA over Converged
Ethernet would be an ideal technology. This technology bypasses copies
in several layers of the TCP/IP stack and uses direct
memory-to-memeory copies to delivers throughputs much closer to the
limits of the network itself. It is available currently for GPU as
GPUdirect (which communicates directly from on GPU's memory to another
over a network), and in Java as Sockets Direct. However, at this time
both these technologies are only available for Infiniband networks.
We will monitor these technologies, and we also plan to experiment
with some open source projects like RoCE (RDMA over Converged
Ethernet) which offer more modest gains.

\begin{figure}
\centering
\includegraphics[scale = 0.5]{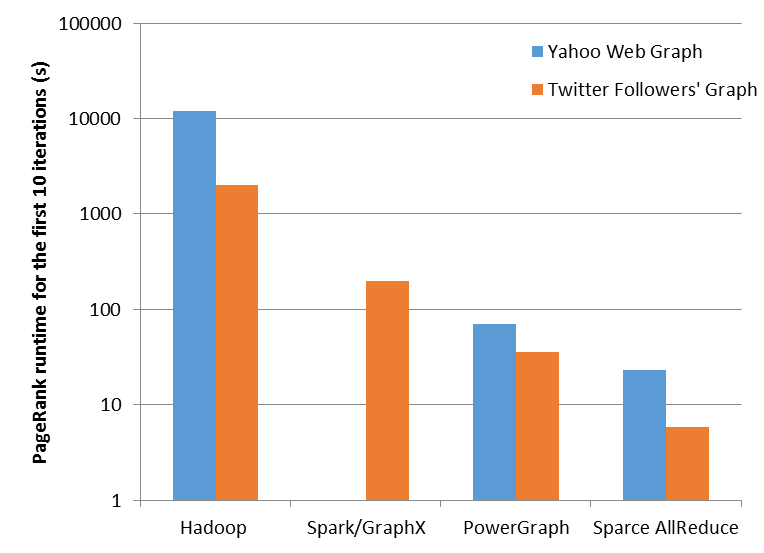}
	\caption{PageRank runtime comparison (log scale).}
\label{benchmark}
\vspace*{-0.2in}
\end{figure} 

Finally, we compare our system with other popular distributed learning
systems: Hadoop/Pegasus, GraphX and GraphLab. Figure \ref{benchmark}
plots the first 10 iteration runtime for different systems. There's no
data available for Mahout and GraphX for the Yahoo dataset. The y-axis
of the plot is log-scale. Sparse Allreduce spends 6 seconds for 10
PageRank iterations on the Twitter followers' graph and 23 seconds for
the Yahoo graph. As seen from the figure, each system provide half to
one order magnitude improvement from right to left.

It also worth mentioning that PowerGraph uses greedily partitioned
graph which produces shorter vertex lists (and communication) on each node.
Our benchmarks use random partitioning, and should improve by about
15-20 \% using greedy partition.

\section{Related Works}
Many other distributed learning systems are under active development
at this time \cite{low2010graphlab, gonzalez12powergraph,
  malewicz2010pregel, ZC13, kang2009pegasus, isard2007dryad}. Our work
is closest to the GraphLab project which also has a focus on Power-Law
graphs and matrices. \cite{low2010graphlab} improves upon the Hadoop
MapReduce infrastructure by expressing asynchronous iterative
algorithms with sparse computational dependencies. PowerGraph is a
improved version of GraphLab, where the concerns about power-law graph
in the context of graph mining has been first proposed. We have taken
a somewhat more modular approach, isolating the Allreduce primitive
from matrix and machine learning modules. The Pegasos project proposed
GIM-V, a primitive generalizable to a variety of graph mining
tasks. We further generalize the primitive to mini-batch update
algorithms which covers regressions, factor model, topic models
etc. There are a variety of other similar distributed data mining
systems \cite{mahout, kang2008hadi, papadimitriou2008disco,
  bu2010haloop} built on top of Hadoop that however, the
disk-caching and disk-buffering philosophy of Hadoop, along with heavy
reliance on reflection and serialization, cause such approaches to fall
orders of magnitude behind the other approaches discussed here. 

Our work is also related with research in distributed SpMV (Sparse
Matrix Vector multiplication) algorithms \cite{demmel2008avoiding,
  hoemmen2010communication, demmel2008communication} in the
parallel/scientific computing community. However, they usually deal
with matrices with regular shapes (tri-diagonal) or desirable partition
properties such as small surface-to-volume ratio. We also distinguish
our work by concentrating on studying the performance trade-off on
commodity hardwares, such as on Amazon EC2, as opposed to scientific
clusters featuring extremely fast network connections, high synchronization
and exclusive (non-virtal) machine use.

\section{Conclusion}
In this paper, we describe Sparse Allreduce for efficient and scalable
distributed machine learning. The primitive is particularly well-adapted
to the power-law data common in machine learning and graph analysis. 
We showed that the best approach is a hybrid between butterfly and
round-robin topologies, using a nested communication pattern and
non-homogeneous layer degrees.
We added a replication layer to the network which provides a high degree
of fault tolerance with modest overhead. 
Finally, we presented a number of experiments exploring the performance
of Sparse Allreduce primitive. We showed that it is significantly 
faster than other primitives, and is limited at this time by the
performance of the underlying technology, Java Sockets. In the future
we hope to achieve further gains by using more advanced network 
layers that use RDMA over Converged Ethernet (RoCE). Our code is
open-source and freely-available, and is currently in pure Java.
It is distributed as part of the BIDMat suite, but can be run standalone
without other BIDMat features. 




\bibliographystyle{IEEEtran}
\bibliography{refs}
%



\end{document}